\documentclass{article}
\usepackage{authblk}
\title{Transfer and evolution of structured polarization in a double-$V$ atomic system}
\author[1]{\footnotesize{Zhenzhu Li}}
\author[2]{Sonja Franke-Arnold}
\author[3]{Thomas W. Clark}
\author[2,4]{Jinwen Wang}
\author[1]{Dawei Zhang}
\author[1]{Chunfang Wang
\thanks{Corresponding author.
E-mail address:cfwang@usst.edu.cn(C.Wang).}}
\affil[1]{Department of Physics, University of Shanghai for Science and Technology, Shanghai 200093, China}
\affil[2]{School of Physics and Astronomy, University of Glasgow, G12 8QQ, United Kingdom}
\affil[3]{Wigner Research Centre for Physics, Hungarian Academy of Sciences, H-1525, Hungary}
\affil[4]{Ministry of Education Key Laboratory for Nonequilibrium Synthesis and Modulation of Condensed Matter, Shaanxi Province Key Laboratory of Quantum Information and Quantum Optoelectronic Devices, School of Physics, Xi'an Jiaotong University, Xi'an 710049, China}
\date{}
\usepackage{amsmath}
\usepackage{graphicx}
\usepackage{subfigure}
\begin{document}
\maketitle
\begin{abstract}
We numerically investigate the transfer of optical information from a vector-vortex control beam to an unstructured probe beam, as mediated by an atomic vapour. The right and left circular components of these beams drive the atomic transitions of a double-$V$ system, with the atoms acting as a spatially varying circular birefringent medium. Modelling the propagation of the light fields, we find that, for short distances, the vectorial light structure is transferred from the control field to the probe. However, for larger propagation lengths, diffraction causes the circular components of the probe field to spatially separate. We model this system for the D1 line of cold rubidium atoms. Our investigation is a first step to investigating the coupled dynamics of internal and external degrees of freedom of atoms in four wave mixing.
\end{abstract}

\section{Introduction}
In the past few decades, the interaction of coherent light with multilevel atomic systems has revealed many unexpected phenomena in nonlinear and quantum optics [1-10]. Of particular importance, was the discovery of electromagnetically induced transparency (EIT) [1,2], where the optical response of an atomic transition is dramatically altered by additional coupling to a third state. EIT is essentially the result of destructive quantum interference between these transitions, such that absorption is severely reduced for a given range of probe frequencies and eliminated entirely on resonance. Such effects are also accompanied by a sharp change in refractive index, yielding applications in all-optical communication [11,12], superluminal and ultraslow light [13,14], light storage [15], giant optical nonlinearities [16,17], optical bistability [18] and multi-wave mixing [19].

Following soon after EIT, was the discovery that light can carry orbital angular momentum (OAM). By imparting an azimuthal phase dependence around a Gaussian beam's axis of propagation, we can ensure that the beam's energy winds around the axis in a helical fashion. The number of windings per wavelength then provides an additional degree of freedom in manipulating optical information, but comes at the cost of dislocations in the phase profile. With strong meteorological analogy, this helical field propagation, and the resulting annular intensity profile, is known as a vortex beam [20].

Since they were first conceived and demonstrated [21,22], OAM-laden beams have found many applications: in communications [23-26], field manipulation [27,28], trapping [29], beam shaping [30] and, lately, a new type of compass [31]. This latter work was based on the marriage of EIT and OAM effects demonstrated by Radwell \textit{et al.} [32]: where coherent interaction of optical OAM and atoms allowed for spatially dependent EIT. Hamedi \textit{et al.} [33] then theoretically investigated the influence of vortex beams on the combined tripod and $\Lambda$ (CTL) atomic medium, showing features of OAM can be transferred to the probe light in a highly resonant medium.

In this paper, we analyze the co-propagation of a vectorial control and a homogeneously polarized probe beam through an extended interaction region.  We model the atomic dynamics in terms of optical Bloch equations and the dynamics of the light fields via split-step propagation methods, and analyze the resulting optical fields in terms of their evolving intensity and polarization profiles on propagation.

In our simulations we find a markedly different behavior depending on the length of the interaction region: For short interaction lengths we show theoretically that vectorial light structure can be transferred from the control beam to the probe beam. For longer interaction regions, however, the probe beam with its imprinted vector structure separates into its left and right circular polarization components, which experience different diffraction, leading to a splitting of the probe beam intensity. We explain this by a spatially varying circuclar birefringence of the atomic medium, induced by the vectorial control beam.

\section{Theoretical model and formulation}

\begin{figure}[ht]
  \centering
  \subfigure
{
    \begin{minipage}[b]{0.3\linewidth}
    \centering
    \includegraphics[scale=0.25]{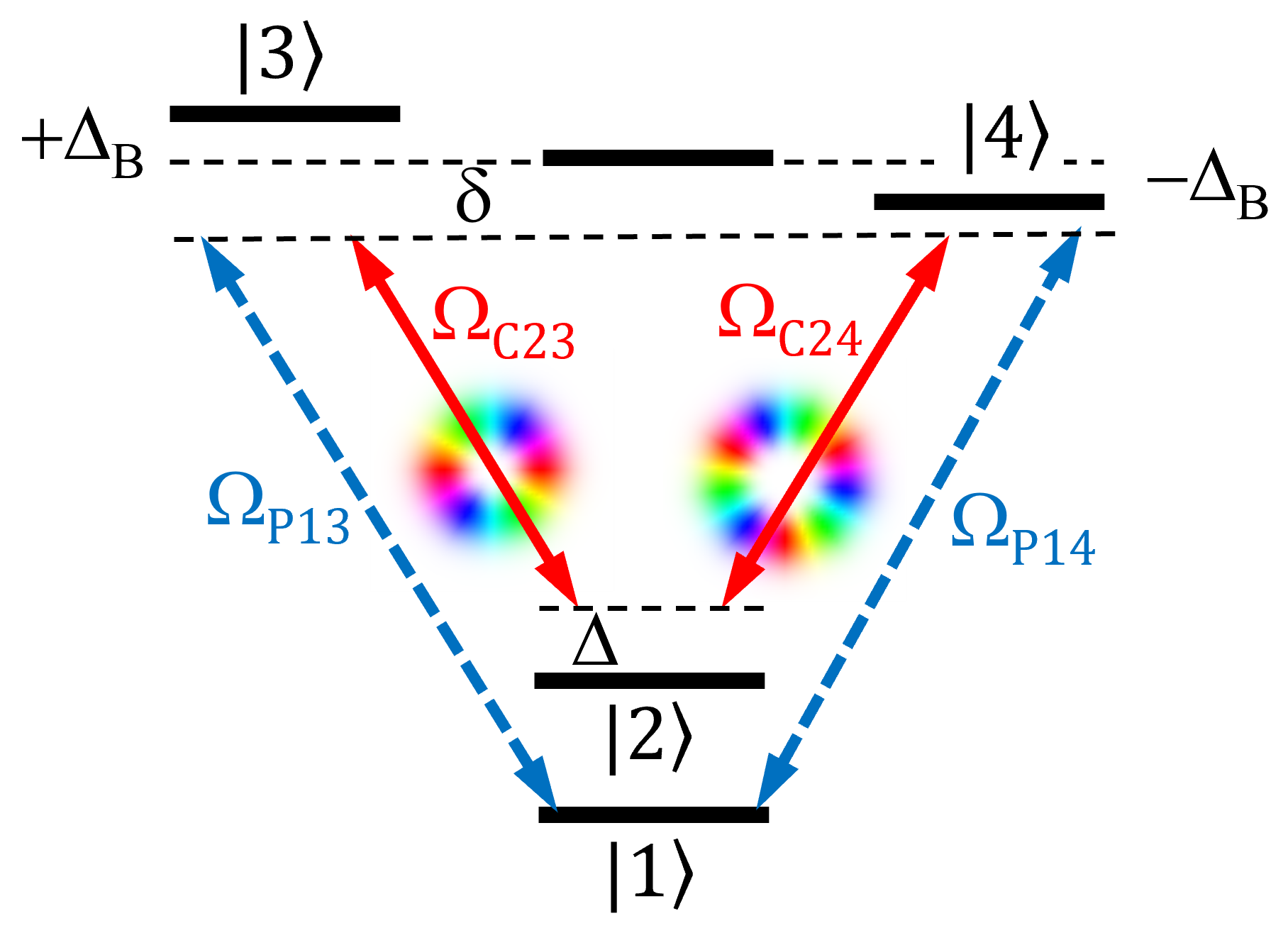}
    \end{minipage}
}
 \subfigure
{
    \begin{minipage}[b]{0.3\linewidth}
    \centering
    \includegraphics[scale=0.3]{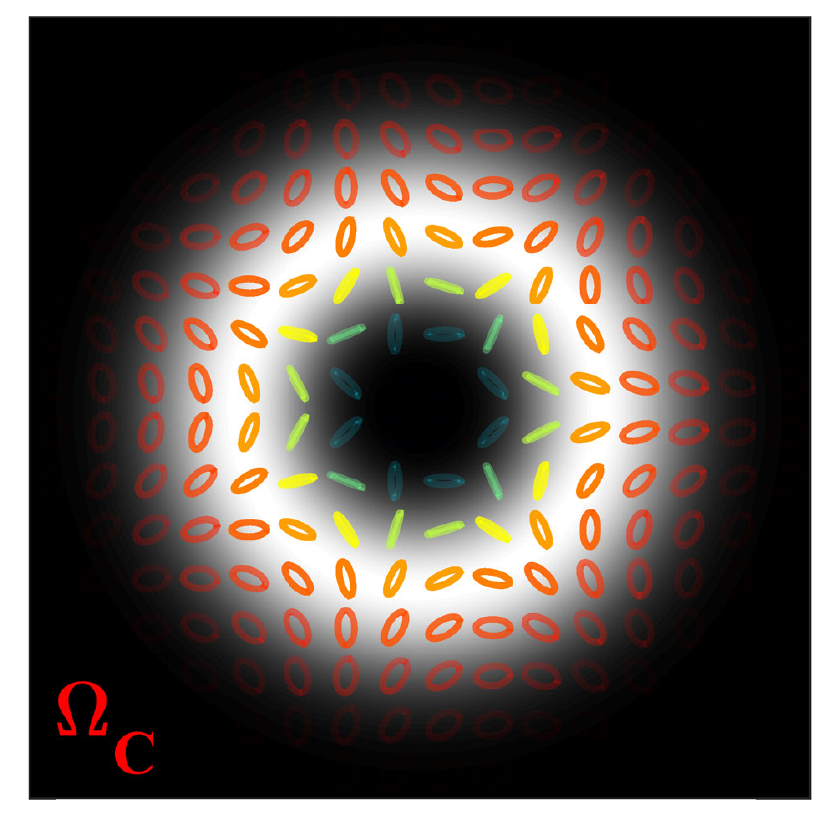}
    \end{minipage}
}
 \subfigure
{
    \begin{minipage}[b]{0.3\linewidth}
    \centering
    \includegraphics[scale=0.3]{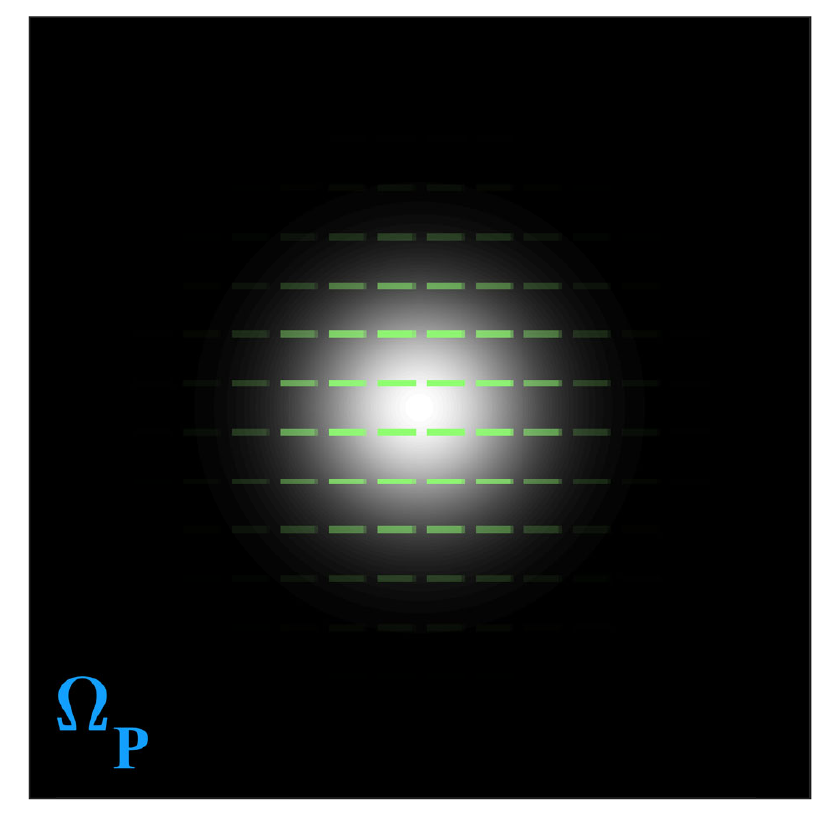}
    \end{minipage}
}
\caption{A schematic diagram of the closed, four-level double-$V$ system. Here, $|1\rangle$ and $|2\rangle$ are two ground states with magnetic quantum number $m_{\rm{F}}=0$. The upper levels, $|3\rangle$ and $|4\rangle$, are magnetic sublevels of the same excited hyperfine state, with $m_{\rm{F}}=-1$ and $+1$ respectively. The right (left) polarization component of the probe couples $|1\rangle$ $\to$ $|3\rangle$ ($|4\rangle$) with a Rabi frequency of $\Omega_{\rm{P13}}$ ($\Omega_{\rm{P14}}$).  Similarly, the right (left) polarization component of the control couples $|2\rangle$ $\to$ $|3\rangle$ ($|4\rangle$) with a Rabi frequency of $\Omega_{\rm{C23}}$ ($\Omega_{\rm{C24}}$). We denote with $\delta$ and $\Delta$ the one-photon detuning and two-photon detuning, respectively, and assuming a magnetic field, the upper states are further shifted by the Zeeman effect $\pm\Delta_{\rm{B}}$. The insets show examples of polarization profiles of the control and probe.}
\label{Fig1}
\end{figure}

We consider classical control and probe beams co-propagating through a dilute atomic gas along the positive z-direction.  The orthogonal circular polarization components of each laser beam couple $V$-type transitions, forming a four-level double-$V$ system (Fig.~\ref{Fig1}).

Such a system can be realized experimentally with $^{87}$Rb: where the atoms are approximated as a non-interacting gas and where for the two ground states, $|1\rangle=|5^{2}S_{1/2},F=1,m_{\rm{F}}=0\rangle$ and $|2\rangle=|5^{2}S_{1/2},F=2,m_{\rm{F}}=0\rangle$, and for upper states, $|3\rangle=|5^{2}P_{1/2},F=1,m_{\rm{F}}=-1\rangle$ and $|4\rangle=|5^{2}P_{1/2},F=1,m_{\rm{F}}=1\rangle$.
A static magnetic field, $\vec{B}=B\hat{z}$, then Zeeman-shifts the levels according to  $\Delta_{\rm{B}}=g_{\rm{F}}\mu_{\rm{B}} B/ \hbar$, where $g_{\rm{F}}$ and $\mu_{\rm{B}}$ denote the Land\'e factor and Bohr magneton, respectively. The atoms interact with a strong vectorial control field, $\vec{E}_{\rm{C}}$, that has a spatially varying polarization profile, and a weak, homogeneously polarized, probe field, $\vec{E}_{\rm{P}}$. We express both laser fields in terms of their circular polarization components, with the right and left handed polarization components coupling to $|3\rangle$ and $|4\rangle$, respectively.

The electric fields of the control and probe beam are given by
\begin{align}
\vec{E}_{\rm{C}}(\vec{r}_\bot,z,t)
& =[\hat{\sigma}_{-}\mathcal{E}_{\rm{C23}}(\vec{r}_\bot,z,t)+
\hat{\sigma}_{+}\mathcal{E}_{\rm{C24}}(\vec{r}_\bot,z,t)]
e^{-i(\omega_{\rm{C}}t-k_{\rm{C}}z)}+c.c.  \text{ and } \nonumber\\
\vec{E}_{\rm{P}}(\vec{r}_\bot,z,t) & =
\mathcal{E}_{\rm{P}}(\vec{r}_\bot,z,t)[\hat{\sigma}_{\rm{+}}+\hat{\sigma}_{\rm{-}}]e^{-i(\omega_{\rm{P}}t-k_{\rm{P}}z)}+c.c., \tag{1},
\end{align}
where $\vec{r}_\bot$, $z$ and $t$ represent the transverse spatial coordinates, the propagation distance within the atomic medium and time, respectively.
Furthermore, $\sigma_{\pm}$ denote the unit vectors of left and right circular polarization, $\omega_{\rm{C}}$ ($\omega_{\rm{P}}$) and $k_{\rm{C}}$ ($k_{\rm{P}}$) are the angular frequency and wavenumbers for the control (probe) beam, and $\mathcal{E}_{\rm{C23}}$ and $\mathcal{E}_{\rm{C24}}$ ($\mathcal{E}_{\rm{P}}$) are the slowly varying envelopes of the control (probe) beam. For the control field, with its spatially varying polarization profile, the complex amplitudes of the right and left circular polarization components differ: their local phase difference determines the direction of the major axis of the polarisation ellipses, and their local amplitude difference, the degree of ellipticity.

The light-matter interaction is then characterized by the Rabi frequencies
\begin{align}
\Omega_{\rm{C23}}(\vec{r}_\bot,z,t) &=\frac{\vec{d}_{23}\cdot
\hat{\sigma}_{\rm{-}}}{\hbar}\mathcal{E}_{\rm{C23}}(\vec{r}_\bot,z,t), \quad
& \Omega_{\rm{C24}}(\vec{r}_\bot,z,t)  =\frac{\vec{d}_{24}\cdot
\hat{\sigma}_{\rm{+}}}{\hbar}\mathcal{E}_{\rm{C24}}(\vec{r}_\bot,z,t),
\nonumber \\
 \Omega_{\rm{P13}}(\vec{r}_\bot,z,t) & =\frac{\vec{d}_{13}\cdot
\hat{\sigma}_{\rm{-}}}{\hbar}\mathcal{E}_{\rm{P}}(\vec{r}_\bot,z,t), \quad
& \Omega_{\rm{P14}}(\vec{r}_\bot,z,t)  =\frac{\vec{d}_{14}\cdot
\hat{\sigma}_{\rm{+}}}{\hbar}\mathcal{E}_{\rm{P}}(\vec{r}_\bot,z,t), \hfill
\tag{2}
\end{align}
where $\vec{d}_{ij}=qr$ are the dipole moments and $r$ is defined as the displacement from $|i\rangle$ to $|j\rangle$.

Note that the transition paths $|3\rangle\to|2\rangle \to |4\rangle \to |1\rangle \to |3\rangle $ constitute a closed-loop double-$V$ coherent coupling system.

Under the rotating-wave approximation, the time-independent Hamiltonian takes the form:
\begin{flalign}
H=&-\hbar[\Delta|2\rangle\langle2|
-(\delta+\Delta_{\rm{B}})|3\rangle\langle3|-
(\delta-\Delta_{\rm{B}})|4\rangle\langle4|]
\notag
\\&-\hbar(\Omega_{\rm{P13}}|1\rangle\langle3|
+\Omega_{\rm{P14}}|1\rangle\langle4|+\Omega_{\rm{C23}}
|2\rangle\langle3|+\Omega_{\rm{C24}}|2\rangle\langle4|
+H.c.),
\tag{3}
\end{flalign}
where $\delta=\omega_{31}-\Delta_{\rm{B}}-\omega_{\rm{P}}$ and $\Delta=\omega_{\rm{P}}-\omega_{\rm{C}}-\omega_{21}$ are the one-photon resonance detuning and  two-photon detuning, respectively.
The atomic dynamics are described by the Liouville equation:
\begin{equation}
\dot{\rho}=-\frac{i}{\hbar}[H,\rho]+\mathcal{L}_{\rm{r}}\rho+\mathcal{L}_{\rm{c}}\rho,
\tag{4}
\end{equation}
where the second term characterizes radiative and the third  non-radiative processes. These can be expressed as
\begin{align}
\mathcal{L}_{\rm{r}}\rho &=-\sum^{4}_{i=3}\sum^{2}_{j=1}
\frac{\gamma_{ij}}{2}(|i\rangle\langle i|\rho-2|j\rangle\langle j|\rho_{ii}+\rho|i\rangle\langle i|),
\tag{5a}\\
\mathcal{L}_{\rm{c}}\rho &=-\sum^{2}_{i=1}
\sum^{2}_{i\neq{j}=1}\frac{\gamma_{\rm{c}}}{2}
(|i\rangle\langle i|\rho-2|j\rangle\langle j|\rho_{ii}+\rho|i\rangle\langle i|).
\tag{5b}
\end{align}
Here $\gamma_{ij}$ represent radiative decay rates from excited states $|i\rangle$ to ground states $|j\rangle$, and $\gamma_{\rm{c}}$ is the collision rate. The dynamics of the atomic population and coherence of the closed-loop system can be obtained by substituting the Hamiltonian Eq.(3) and radiative term Eq.(5) into the Liouville equation Eq.(4), so that the dynamics of the density matrix elements are determined by the optical Bloch equations:

\begin{align}
\dot{\rho}_{11} &=i(\Omega_{\rm{P13}}\rho_{31}-
\Omega^{\ast}_{\rm{P13}}\rho_{13}+\Omega_{\rm{P14}}\rho_{41}-\Omega^{\ast}_{\rm{P14}}\rho_{14})+\gamma_{31}\rho_{33}+
\gamma_{41}\rho_{44}+\gamma_{\rm{c}}(\rho_{22}-
\rho_{11}),
\nonumber\\
\dot{\rho}_{22} &=i(\Omega_{\rm{C23}}\rho_{32}-
\Omega^{\ast}_{\rm{C23}}\rho_{23}+\Omega_{\rm{C24}}\rho_{42}-\Omega^{\ast}_{\rm{C24}}\rho_{24})+\gamma_{32}\rho_{33}+
\gamma_{42}\rho_{44}+\gamma_{\rm{c}}(\rho_{11}-
\rho_{22}),
\nonumber\\
\dot{\rho}_{33} &=i(\Omega^{\ast}_{\rm{P13}}\rho_{13}-
\Omega_{\rm{P13}}\rho_{31}+\Omega^{\ast}_{\rm{C23}}\rho_{23}-\Omega_{\rm{C23}}\rho_{32})-(\gamma_{31}+
\gamma_{32})\rho_{33},
\nonumber
\\
\dot{\rho}_{12} &=i[\Omega_{\rm{P13}}\rho_{32}-\Omega^{\ast}_{\rm{C23}}\rho_{13}-\Omega^{\ast}_{\rm{C24}}\rho_{14}+\Omega_{\rm{P14}}\rho_{42}-\Delta
\rho_{12}]-\gamma_{c}\rho_{12},
\nonumber
\\
\indent
\dot{\rho}_{13} &=i[\Omega_{\rm{P13}}(\rho_{33}-\rho_{11})-\Omega_{\rm{C23}}\rho_{12}+
\Omega_{\rm{P14}}\rho_{43}+(\delta+\Delta_{\rm{B}})\rho_{13}]
- \Gamma_{31}\rho_{13},
\nonumber
\\
\dot{\rho}_{14} &=i[\Omega_{\rm{P14}}(\rho_{44}-\rho_{11})-\Omega_{\rm{C24}}\rho_{12}+
\Omega_{\rm{P13}}\rho_{34}+(\delta-\Delta_{\rm{B}})\rho_{14}]-\Gamma_{41}\rho_{14},
\nonumber
\\
\dot{\rho}_{23} &=i[\Omega_{\rm{C23}}(\rho_{33}-\rho_{22})-\Omega_{\rm{P13}}\rho_{21}+
\Omega_{\rm{C24}}\rho_{43}+(\delta+\Delta_{\rm{B}}+\Delta)\rho_{23}]
-\Gamma_{32}\rho_{23},
\nonumber
\\
\dot{\rho}_{24} &=i[\Omega_{\rm{C24}}(\rho_{44}-\rho_{22})-\Omega_{\rm{P14}}\rho_{21}+
\Omega_{\rm{C23}}\rho_{34}+(\delta-\Delta_{\rm{B}}+\Delta)\rho_{24}]-\Gamma_{42}\rho_{24},
\nonumber
\\
\dot{\rho}_{34} &=i[\Omega^{\ast}_{\rm{P13}}\rho_{14}-\Omega_{\rm{P14}}\rho_{31}-\Omega_{\rm{C24}}\rho_{32}+\Omega^{\ast}_{\rm{C23}}\rho_{24}-2\Delta_{\rm{B}}\rho_{34}]-\frac{\gamma_{31}+\gamma_{32}+\gamma_{41}+\gamma_{42}}{2}\rho_{34},
\nonumber
\\
\dot{\rho}_{44} &=-(\dot{\rho}_{11}+\dot{\rho}_{22}
+\dot{\rho}_{33}),
\tag{6}
\end{align}
and $\dot{\rho}_{ji}=\dot{\rho}^{\ast}_{ij}$. The coherence decay terms are defined as $\Gamma_{ij}=\gamma_{\rm c}/2+\gamma_{ij}.$  In the following we assume balanced radiative decay rates of $\gamma_{31}=\gamma_{32}=\gamma_{41}
=\gamma_{42}=\gamma/2$, where $\gamma$ is the spontaneous upper state decay rate, so that the coherence decay terms simplify to $\Gamma_{31}=\Gamma_{32}=\Gamma_{41}=\Gamma_{42}
=(\gamma+\gamma_{\rm{c}})/2$.

The atomic susceptibilities $\chi_{1j}$ denotes the response of the atomic medium to the electric pump field, with the (slowly varying) atomic polarization given by $\vec{P}_{1j}=\epsilon_0 \chi_{1j} \vec{E}+{\rm P}.$ This macroscopic polarization is on the atomic level related to the induced dipole moment, as $\vec{P}_{1j}=\mathcal{N} \langle \vec{d}_{1j} \rho_{1j}\rangle,$  where  $\mathcal{N}$ and $\epsilon_{0}$ are the atomic density and vacuum permittivity, respectively.

In order to derive the atomic susceptibilities, we therefore have to indentify the steady state solution of the corresponding optical Bloch equations Eq.~(6). We assume that our probe fields (but not the control fields) are weak enough to be treated as a perturbation, with $\{ \Omega_{\rm{P13}} / \gamma ,\, \Omega_{\rm{P14}}/\gamma \} \ll 1$.
This allows us to expand the density matrix elements to first order in the probe field Rabi frequencies and to neglect all higher order terms of $\Omega_{\rm{P}}$:

\begin{equation}
\rho_{ij}=\rho^{(0)}_{ij}+\rho^{(1)}_{ij}+ \mathcal{O}^2\left(\frac{\Omega_P}{\gamma}\right),  \tag{7}
\end{equation}

\noindent where $\rho^{(0)}_{ij}$ is the zeroth-order  and $\rho^{(1)}_{ij}$ the first-order solution in the probe field $\Omega_{P}$.
Substituting Eq.~(7) into Eq.~(6), the equations of motion for the first-order density-matrix elements, $\rho_{1j}$, can be expressed as
\begin{align}
\dot{\rho}^{(1)}_{12} &=i[\Omega_{\rm{P13}}\rho^{(0)}_{32}-\Omega^{\ast}_{\rm{C23}}\rho^{(1)}_{13}-\Omega^{\ast}_{\rm{C24}}\rho^{(1)}_{14}+\Omega_{\rm{P14}}\rho^{(0)}_{42}-\Delta
\rho^{(1)}_{12}]-\gamma_{\rm{C}}\rho^{(1)}_{12},
\tag{8a}
\\
\indent
\dot{\rho}^{(1)}_{13} &=i[\Omega_{\rm{P13}}(\rho^{(0)}_{33}-\rho^{(0)}_{11})-\Omega_{\rm{C23}}\rho^{(1)}_{12}+
\Omega_{\rm{P14}}\rho^{(0)}_{43}+(\delta+\Delta_{\rm{B}})\rho^{(1)}_{13}]
- \Gamma_{31}\rho^{(1)}_{13},
\tag{8b}
 \text{and} \\
\dot{\rho}^{(1)}_{14} &=i[\Omega_{\rm{P14}}(\rho^{(0)}_{44}-\rho^{(0)}_{11})-\Omega_{\rm{C24}}\rho^{(1)}_{12}+
\Omega_{\rm{P13}}\rho^{(0)}_{34}+(\delta-\Delta_{\rm{B}})\rho^{(1)}_{14}]-\Gamma_{41}\rho^{(1)}_{14}.
\tag{8c}
\end{align}

We can assume that the atomic population is initially distributed between the ground states $|1\rangle$ and $|2\rangle$, so that to zeroth-order, $\rho_{11}^{(0)}+\rho_{22}^{(0)}=1$, with all other elements $\rho_{ij}^{(0)}=0$. Ignoring higher-order components, the steady state solutions are then

\begin{align}
\rho^{(1)}_{13} & =\frac{i(\xi_{1}\xi_{3}\Omega_{\rm{P13}}
+|\Omega_{\rm{C24}}|^2\Omega_{\rm{P13}}-
\Omega_{\rm{P14}}\Omega_{\rm{C24}}^{\ast}\Omega_{\rm{C23}})}
{2(\xi_{1}\xi_{2}\xi_{3}+\xi_{3}|\Omega_{\rm{C23}}
|^2+\xi_{2}|\Omega_{\rm{C24}}|^2)},
\tag{9a}\\
\rho^{(1)}_{14} & =\frac{i(\xi_{1}\xi_{2}\Omega_{\rm{P14}}
+|\Omega_{\rm{C23}}|^2\Omega_{\rm{P14}}-
\Omega_{\rm{P13}}\Omega^{\ast}_{\rm{C23}}\Omega_{\rm{C24}})}
{2(\xi_{1}\xi_{2}\xi_{3}+\xi_{3} |\Omega_{\rm{C23}}
|^2+\xi_{2}|\Omega_{\rm{C24}}
|^2},
\tag{9b}
\end{align}
\noindent where we have defined $\xi_{1}=-i\Delta-\gamma_{\rm{C}}$,
$\xi_{2}=i(\delta+\Delta_{\rm{B}})-\Gamma_{31}$,and
$\xi_{3}=i(\delta-\Delta_{\rm{B}})-\Gamma_{41}$.
Hence, the susceptibilities of the medium can be expressed as

\begin{align}
\chi_{13}=& \frac{\mathcal{N}|\vec{d}_{13}|^2}
{\epsilon_{0}\hbar\Omega_{\rm{P13}}}\rho^{(1)}_{13} 
= \frac{i\mathcal{N}|\vec{d}_{13}|^2(\xi_{1}\xi_{3}
+|\Omega_{\rm{C24}}|^2-
\Omega_{\rm{P14}}\Omega^{\ast}_{\rm{C24}}\Omega_{\rm{C23}}/\Omega_{\rm{P13}})}
{2\epsilon_{0}\hbar(\xi_{1}\xi_{2}\xi_{3}+\xi_{3}|\Omega_{\rm{C23}}|^2+\xi_{2}|\Omega_{\rm{C24}}|^2},\tag{10a}
\\
\chi_{14}=& \frac{\mathcal{N}|\vec{d}_{14}|^2}
{\epsilon_{0}\hbar\Omega_{\rm{P14}}}\rho^{(1)}_{14} 
= \frac{i\mathcal{N}|\vec{d}_{14}|^2(\xi_{1}\xi_{2}
+|\Omega_{\rm{C23}}|^2-\Omega_{\rm{P13}}
\Omega^{\ast}_{\rm{C23}}\Omega_{\rm{C24}}/\Omega_{\rm{P14}})}
{2\epsilon_{0}\hbar(\xi_{1}\xi_{2}\xi_{3}+\xi_{3}|\Omega_{\rm{C23}}|^2+\xi_{2}|\Omega_{\rm{C24}}|^2)},
\tag{10b}
\end{align}

\noindent The imaginary and real parts of $\chi_{13}$ and $\chi_{14}$ represent the absorption and dispersion for the right and left circular polarization components of the probe field, characterizing the circular dichroism and birefringence of the atomic medium. If the complex light amplitudes vary across the beam profile, so do the Rabi-frequencies and the resulting optical activity. The susceptibility of the medium, as experienced by the probe beam, is shaped by the spatial polarization profile of the control beam. Of particular relevance is the final term in the numerator which describes the interference between the two excitation amplitudes driven by the control beam.

In the following section, we illustrate this for the example of a fundamental Gaussian probe beam and a control beam that is composed of different Laguerre-Gaussian modes in its circular polarization components. According to Eq.~(2), the Rabi frequencies are proportional to the electric field envelopes, which now take the form
\begin{align}
\Omega_{\rm{n}}(r,\phi,z)
= & \Omega_{\rm{n0}}\frac{w_{\rm{n}}}{w(z)}\left(\frac{r\sqrt{2}}{w(z)}\right)^{|l_{\rm{n}}|}\exp \left( -\frac{r^{2}}{w^{2}(z)} \right)
L^{|l_{\rm{n}}|}_{p_{\rm{n}}} \left(\frac{2r^{2}}{w^{2}(z)} \right)\exp \left( il_{\rm{n}}\phi \right)
\notag
\\& \times \exp \left( -\frac{ik_{\rm{n}}r^2}{2R(z)} \right)
\exp \left( i(2p_{\rm{n}}+|l_{\rm{n}}|+1)\tan^{-1}(z/z_{\rm{R}}) \right),
\tag{11}
\end{align}
where $n \in {P13, P14, C23, C24}$.
The terms $\Omega_{\rm{n0}}$ for ($\rm{n}\in \{C23,C24,P\}$) combine the amplitude of the light field and a proportionality factor arising from the relevant dipole operator of the atomic transition.  The parameters $p_{\rm{n}}$ and $l_{\rm{n}}$ describe the radial index and topological charge of the LG fields.  The radius of curvature and the Rayleigh length are defined as $R(z)=z+(z^{2}_{\rm{R}}/z)$ and $z_{\rm{R}}=\pi w^{2}_{\rm{n}}/\lambda_{\rm{n}}$, respectively, where $w_{\rm{n}}$ is the beam waist at z=0. The beam width varies with propagation distance, z, as $w(z)=w_{\rm{n}}\sqrt{1+(z/z_{\rm{R}})^{2}}$, and $\phi=\tan^{-1}{(y/x)}$ is the azimuthal angle. Light beams of this form (Eq.~(11)) carry an orbital angular momentum (OAM) of $l_{\rm{n}} \hbar$ per photon and for a homogeneously linearly polarized probe beam with $|\vec{d}_{13}|=|\vec{d}_{14}|$, we further simplify Eqs.~(10): $\Omega_{\rm{P13}}=\Omega_{\rm{P14}}=\Omega_{\rm{P}},$.

The OAM carried in the control beams then plays an important role in subsequent formation of structured light. Following the simplification above, the final term in the numerator of the susceptibility is $\Omega_{\rm{C23}}\Omega_{\rm{C24}}^{\ast}\propto \exp[i (l_{\rm C23}-l_{\rm C24})]$ for $\chi_{13}$ and its complex conjugate for $\chi_{14}$, which can be interpreted as quantum interference. The difference in OAM between the two orthognal polarization components of the control beam (manifest as a spatially varying polarization) therefore leads to a spatially varying atomic susceptibility which in turn modifies the orthogonal polarization components of the probe beam upon propagation.

In order to study this effect of the control field on the probe beam, we evaluate Maxwell's wave equations to describe the dynamic behavior of the probe field. We assume slowly varying envelopes and work within the paraxial wave approximation. The dynamics along the z-direction can then be expressed by
\begin{equation}
\frac{\partial\Omega_{\rm{P13}}}{\partial z}=\frac{i}{2k_{\rm{P}}}\left(\frac{\partial^2}
{\partial x^2}+\frac{\partial^2}
{\partial y^2}\right)\Omega_{\rm{P13}}+\frac{ik_{\rm{P}}\chi_{13}}{2}\Omega_{\rm{P13}},
\tag{12a}\\
\end{equation}

\begin{equation}
\frac{\partial\Omega_{\rm{P14}}}{\partial z}=\frac{i}{2k_{\rm{P}}}\left(\frac{\partial^2}
{\partial x^2}+\frac{\partial^2}
{\partial y^2}\right)\Omega_{\rm{P14}}+\frac{ik_{\rm{P}}\chi_{14}}{2}\Omega_{\rm{P14}}
\tag{12b},
\end{equation}
\noindent where the first terms on the right-hand side of Eq.~(12) account for diffraction of the probe beam in the medium, while the second terms are responsible for absorption and dispersion. It should be noted that the diffraction terms can only be ignored when the propagation distance is much smaller than the Rayleigh length,  \textit{i.e.} $L\ll\pi w^{2}_{\rm{P}}/\lambda_{\rm{P}}$. In the following, however, we assume a situation where diffraction is not neglible.

\section{Results and discussions}

We have seen that the induced susceptibility, Eq.(10), plays a vital role in describing the properties of the atomic medium. In this section we model the susceptibilities and resulting intensity and polarization profiles of the probe beam for various examples of control beam structures.

To be specific, we consider a propagation distance of up to $L=7$mm (given by the length of the atomic sample), with a beam waist of $w_{\rm{P}}=30\mu$m, and probe wavelength of $\lambda_{\rm{P}}=795$nm, such that $z_{\rm{R}}=\pi
w^{2}_{\rm{P}}/\lambda_{\rm{P}}=3.56$ mm, \textit{i.e.} in a regime where diffraction effects become important. We assume a  $45\times45\mu$m span of the atomic cloud.

Fig.~2 shows the imaginary and real parts of the susceptibilities for control beams with equal intensities in the orthogonal polarization components so that $|\Omega_{\rm C23}|=|\Omega_{\rm C24}|$, but carrying opposite topological charges of $\pm 1, \pm 2$ and $\pm 3$ in rows (a) to (c). The first and second columns are the imaginary parts of $\chi_{13}$ and $\chi_{14}$, while  the third and forth columns are the real parts, corresponding to the absorption and dispersion of the two circular polarization components of the probe field, respectively. The spatially dependent absorption and dispersion show petal-like patterns with petal number $2|l_{\rm{C23}}-l_{\rm{C24}}|$. This is a direct consequence of the  factor $\exp[\pm i(l_{\rm{C23}}-l_{\rm{C24}})\phi]$ in Eqs.~(10-12), which at specific azimuthal positions renders an opaque medium transparent, so that the orthogonal components of the probe field can pass through the medium, almost without loss. Therefore, the periodic oscillations in the susceptibility are responsible for the creation of structured light. We notice that the probe components' absorption profiles are identical, whereas their dispersion profiles are mirror opposites.

 \begin{figure}[ht]
\centering
\subfigure
{
    \begin{minipage}[b]{1\linewidth}
    \centering
    \includegraphics[scale=0.5]{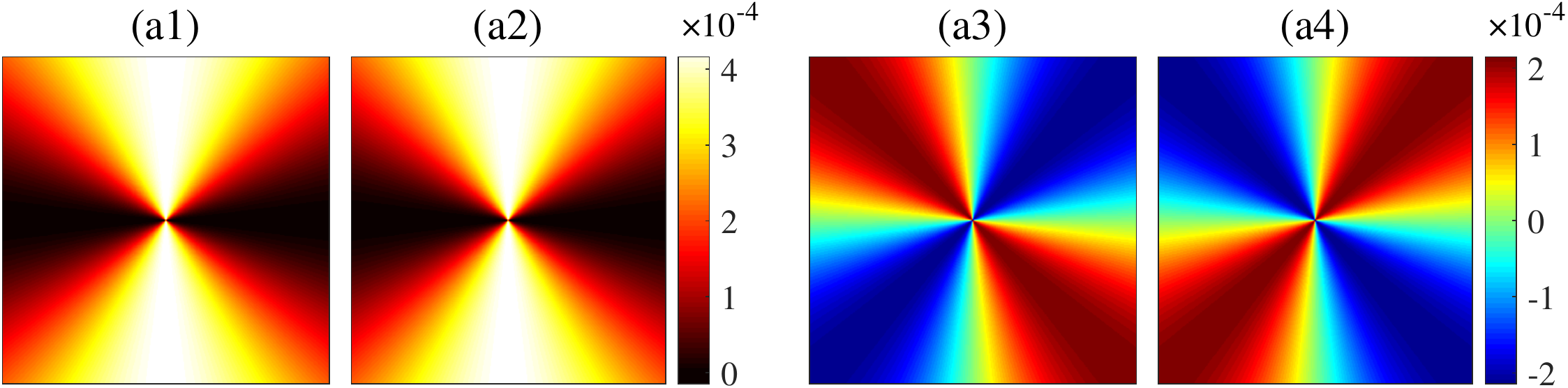}
    \end{minipage}
}

\subfigure
{
    \begin{minipage}[b]{1\linewidth}
    \centering
    \includegraphics[scale=0.5]{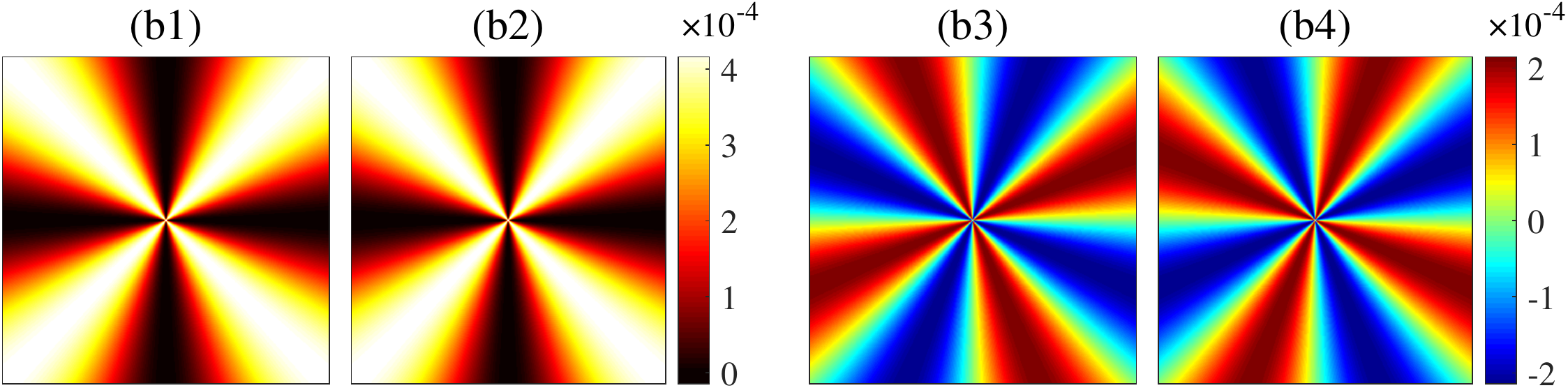}
    \end{minipage}
}

\subfigure
{
    \begin{minipage}[b]{1\linewidth}
    \centering
    \includegraphics[scale=0.5]{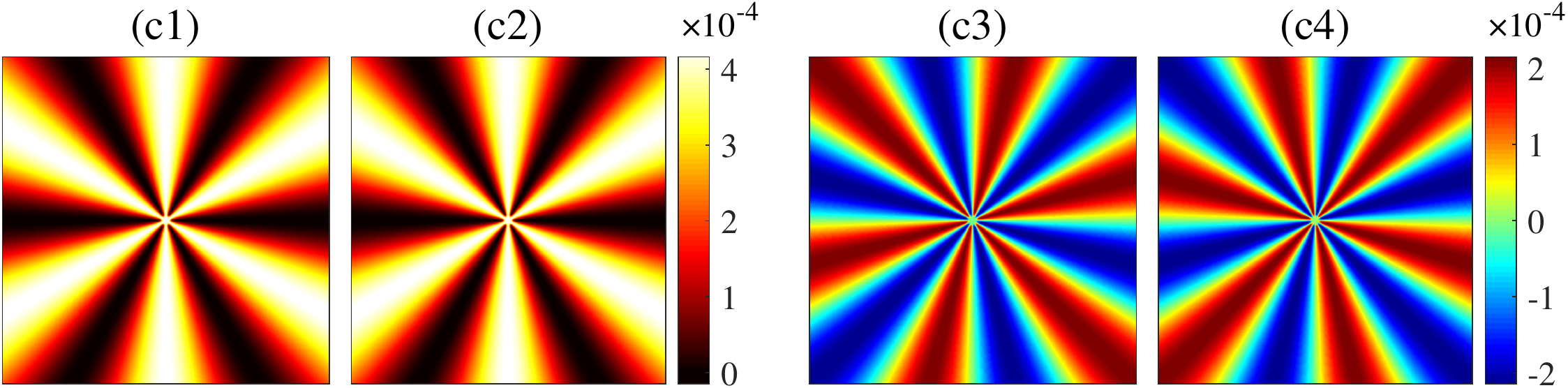}
    \end{minipage}
}
\caption{Linear susceptibility for control beams with different polarization structures. The first and second column are the imaginary parts of $\chi_{13}$ and $\chi_{14}$, and the third and forth columns the corresponding real parts. The different rows show the effect of control beams with different OAM: (a) $l_{\rm{C23}}=-l_{\rm{C24}}=1$, (b) $l_{\rm{C23}}=-l_{\rm{C24}}=2$, (c) $l_{\rm{C23}}=-l_{\rm{C24}}=3$.
In order to satisfy the perturbation condition, we choose $\mathcal{N}=2\times 10^{10}$ atoms/cm$^{3}$, $L=7$mm, $w_{\rm{P}}=30\mu$m, $l_{\rm{P13}}=l_{\rm{P14}}=0$, and $p=0$ for all control and probe beams,
$\Omega_{\rm{P130}}=\Omega_{\rm{P140}}=0.01\gamma$,
$\Omega_{\rm{C230}}=\Omega_{\rm{C240}}=\gamma$, $\delta=\Delta=0$, $\Delta_{\rm{B}}=0.01\gamma$, $\gamma_{\rm{C}}=10^{-7}\gamma$, where $\gamma$=36.1285MHz.}
\end{figure}
We then used the symmetric split-step Fourier method to calculate the intensity of the two orthogonal probe components in different transverse planes. Fig.~3 shows the resulting superimposed intensities, $I_{\rm{out}}\propto|\Omega_{\rm{P13}}|^{2} + |\Omega_{\rm{P14}}|^{2}$, for a control beam with topological charges $l_{\rm{C23}}=-2$ and $l_{\rm{C24}}=2$, as discussed in row (b) of Fig.2. At short propagation distances we observe a four-fold petal structure, however, as the transmission distance increases each petal splits into two parts.

 \begin{figure}[ht]
\centering
\center{\includegraphics[scale=0.6]{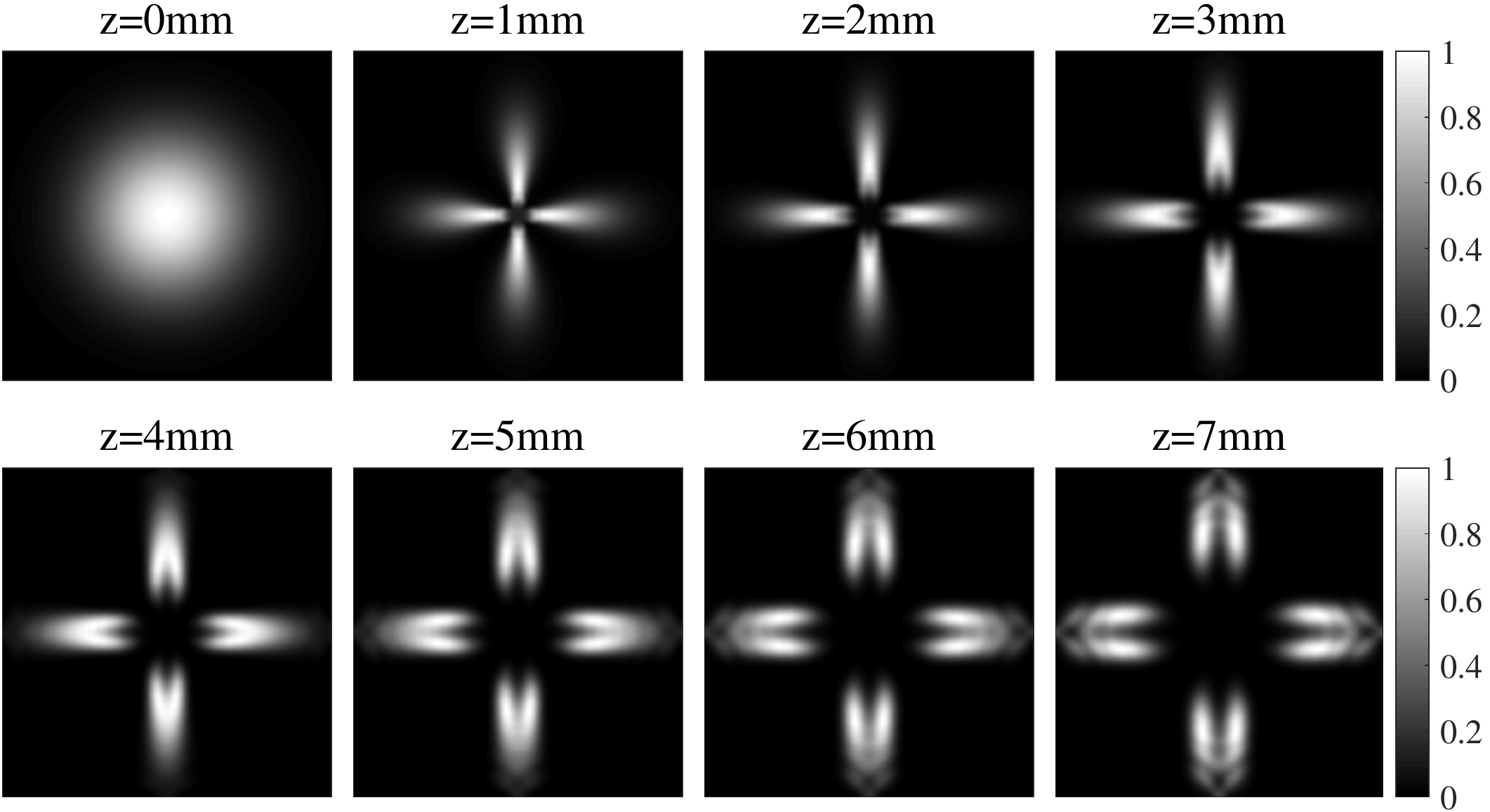}}
\caption{Normalized transmission intensity distribution of the probe beam on different transverse plane with $l_{\rm{C23}}=-2$, $l_{\rm{C24}}=2$, corresponding to the linear susceptibility in Fig.~2(b). Other parameters are the same as those in Fig.2.}
\end{figure}

The four-fold pattern can be explained by considering the absorption profile, which according to Eq.~(10) and Fig.~2(b1) and (b2) has the same effect on the two circular polarization components of the probe field. To explain the splitting  phenomenon however, we must consider dispersion, which differs for the two circular polarizations as shown in Fig.~2(b3) and (b4). This becomes apparent when we depict the individual transmitted intensity distributions of the circular components of the probe field at $z=7$mm in Fig.4. The different dispersion profiles result in differing refractive indices, $n=\sqrt{1+{\rm Re}(\chi)}\approx 1+{\rm Re}(\chi)/2$, which must lead to differing diffraction. The medium can be regarded as an azimuthally-varying diffractive element such that the components split and interfere spatially. With increasing propagation distance, especially under the condition of $z>z_{\rm{R}}$, the effect is more apparent, as shown in Fig.3. The resulting profiles are thus clearly affected by both the absorption and dispersion.

 \begin{figure}[htbp]
\centering
\center{\includegraphics[scale=0.45]{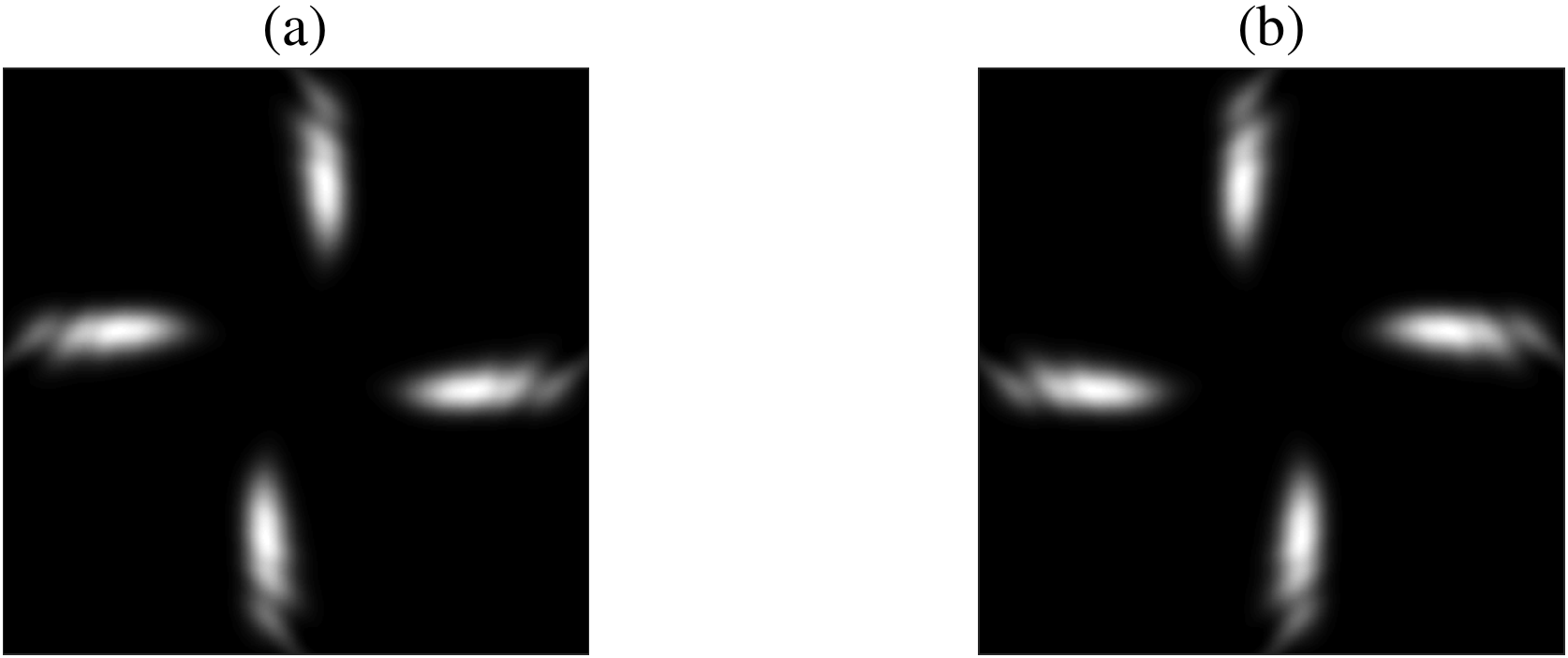}}

\caption{Normalized transmission intensity distribution of the probe beam at $z=7$mm. (a) right-handed circular polarized component; (b) left-handed circular polarized component. Parameters are the same as those in Fig.3}
\end{figure}

In order to give a more comprehensive description of the transmitted probe field, here we investigate the dynamic polarization.  Suppose the probe is initially incident at $z=0$ and linearly polarized in the x-direction. In Fig.5, we then see that as the propagation distance increases, the polarization of the probe field experiences spatially dependent rotation, which  dynamically varies along both angular and radial coordinates. The probe field is then vectorized.

\begin{figure}[h!]
  \centering
  \subfigure{
    \includegraphics[scale=0.6]{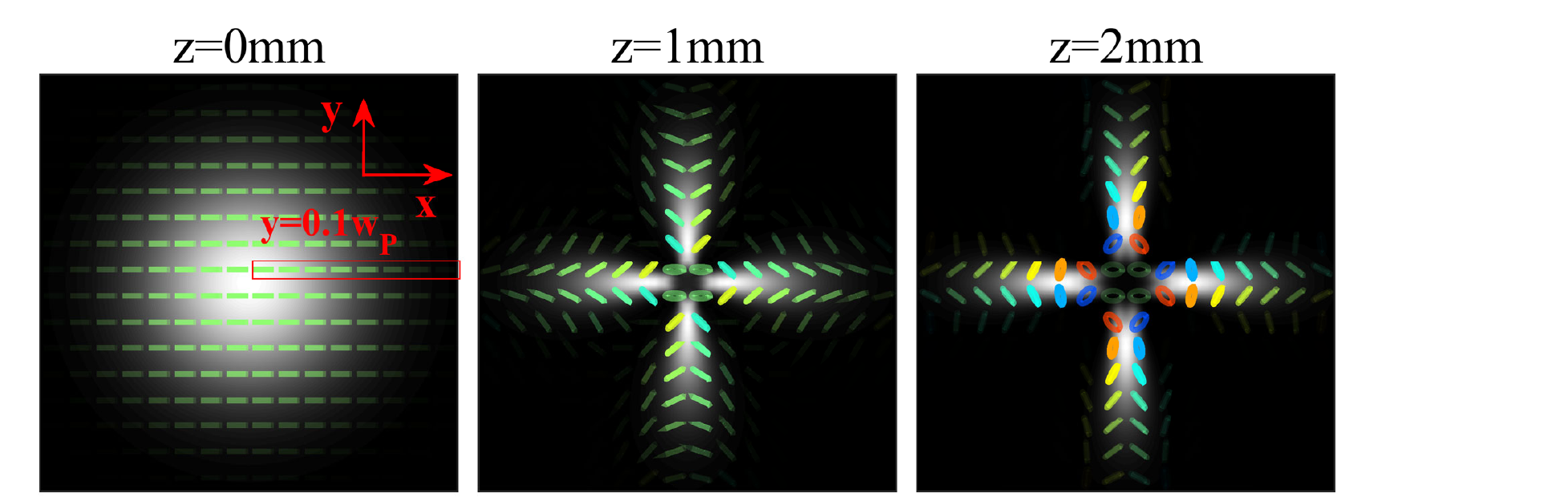}}
  \subfigure{
    \includegraphics[scale=0.6]{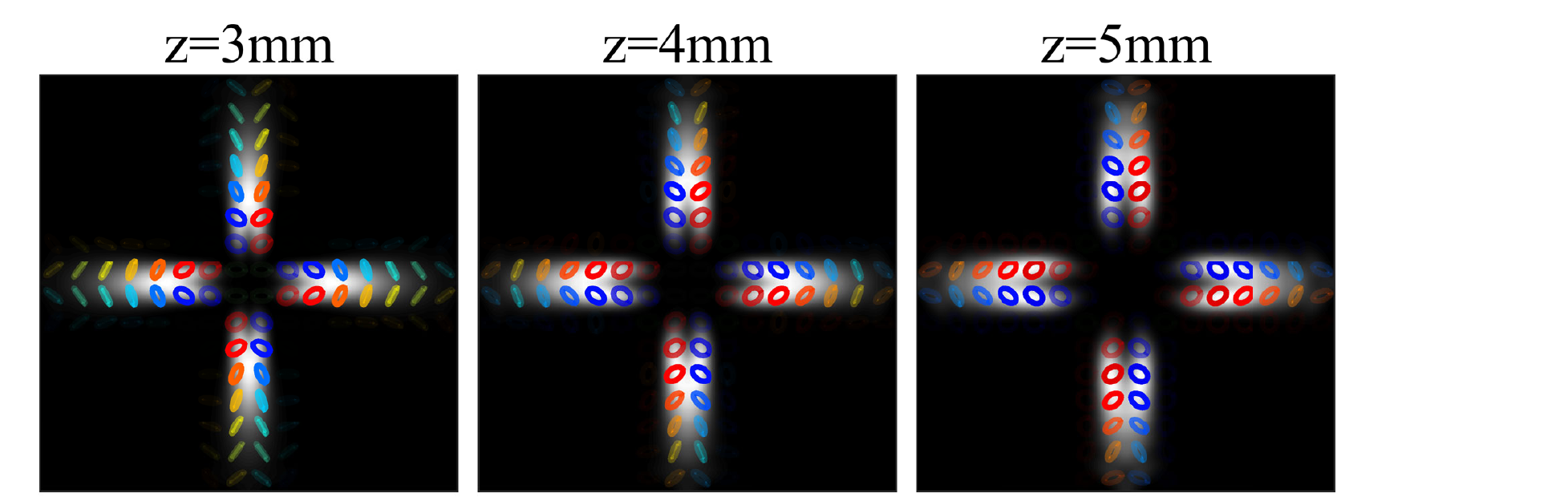}}
 \subfigure{
    \includegraphics[scale=0.6]{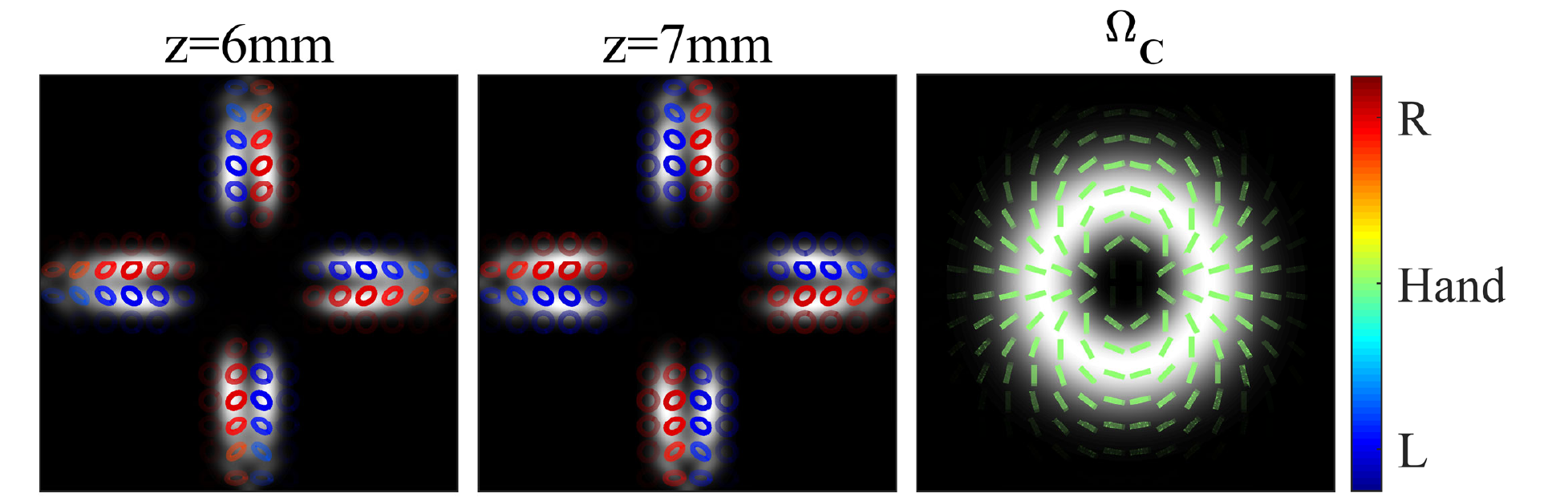}}
\caption{Analysis of the dynamic polarization of the probe field,  with the polarization of control field shown in bottom-right corner. Other parameters are the same as Fig.3.}
\end{figure}

We have seen from Fig.~2 that the different topological charges of the control fields, $\vec{E}_{\rm C24}$ and $\vec{E}_{\rm C24}$, induce different refractive indices for each polarization component: suggesting light-induced, spatially dependent, circular birefringence. The atomic medium thus acts as an optical rotation channel that structures the probe light on interaction. One may expect a Faraday rotation of
$\theta \propto z[Re(\chi_{14})-Re(\chi_{13})]$.
Fig.~5 shows that for our atomic medium, this linear increase of rotation angle with distance holds only for very short propagation distances, and moreover is not uniform across the beam diameter. For larger distances, the change of the rotation angle, $\theta$, is decreasing, and the linear input polarization of the probe beam gradually separates into areas of right and left circular polarisation. One may be tempted to interpret the emerging increase in ellipticity as an effect of circular dichroism, \textit{i.e.}~a differential absorption of left- and right-handed light, but a more accurate interpretation is a differential diffraction of left- and right-handed light.
At larger propagation distances, this diffraction becomes too large to accurately `track' the polarization rotation, an effect that becomes more pronounced for higher-order structured light [34].

To get a better understanding of how optical rotation behaves with propagation, we plot the rotation angle and ellipticity of the probe beam as a function of z (Fig.~6.). Taking advantage of the symmetry, we consider a region just above the horizontal, for $y=0.1w_{\rm{P}}$, $0 \leq x \leq 1.5 w_{\rm{P}}$, and $0 \leq  z \leq 2 z_{\rm R}$, as indicated by the red line in the first panel of Fig.~5. According to Fig.6(a), our simulation shows regions of clockwise (CW) and counter-clockwise (CCW) rotation, indicated in blue and red, respectively. The darker the color, the greater the rotation angle, $\theta$. However, the CCW rotation occurs only close to the optical axis for $|x|<0.1 w_{\rm{P}}$. In this region, the intensity of the probe is extremely low as it is close to the singularity of the optical vortex, and it decreases even further due to the divergence, so that the CCW rotation becomes meaningless. In the other areas of the probe beam, optical rotation, $\theta$, varies with the transverse position, $x$, as well as propagation distance, $z$. The overall trend shows a rotation angle that decreases with distance from the optical axis and increases with propagation distance but saturates for distances larger than the Rayleigh length, corresponding to panels $z=4$mm and above in Fig.~5.

In addition to a rotation of linear polarization, or more generally the major axis of the polarization ellipse, also the degree of ellipticity of the probe itself can change, as evident in Fig.~5. The homogeneous, linear, polarization of the probe beam spatially separates into areas of predominantly right or left polarization for propagation distances beyond the Rayleigh length. As discussed before, the effect is reminiscent of circular dichroism, however its origin lies in circuar `bi-diffraction'.

\begin{figure}[htbp]
\centering
  \subfigure{
    \includegraphics[scale=0.45]{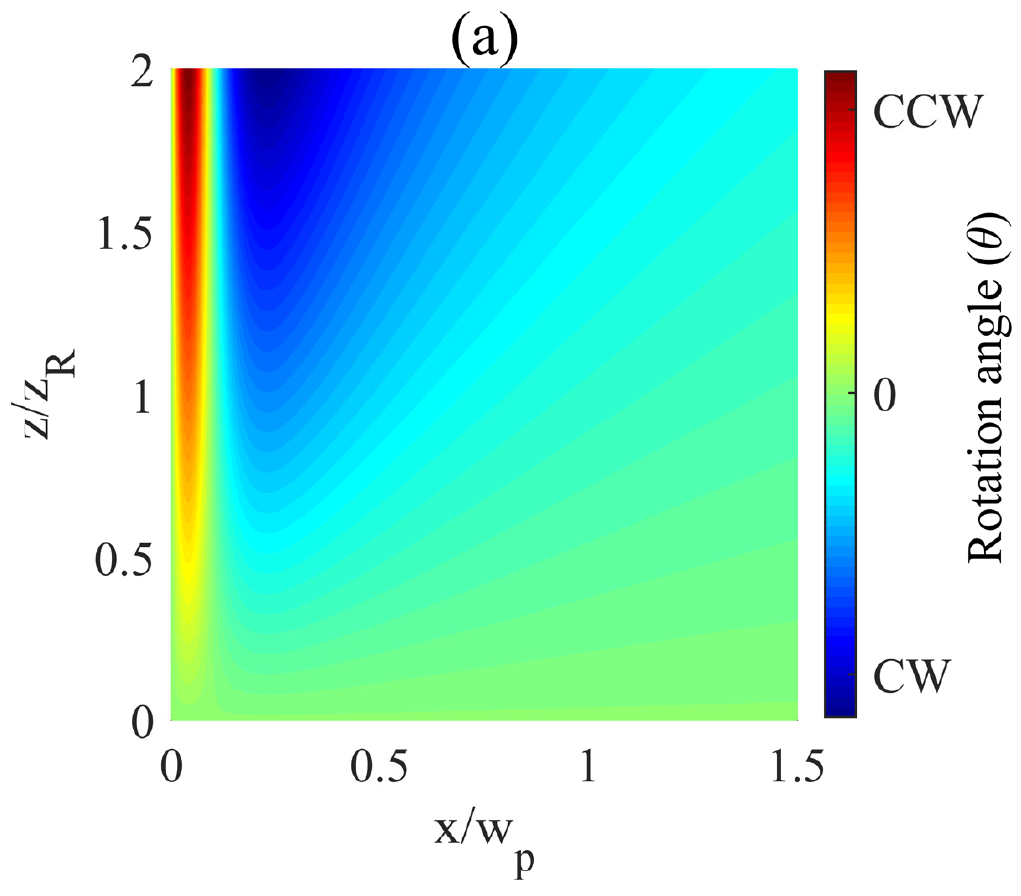}}
    \quad
  \subfigure{
    \includegraphics[scale=0.45]{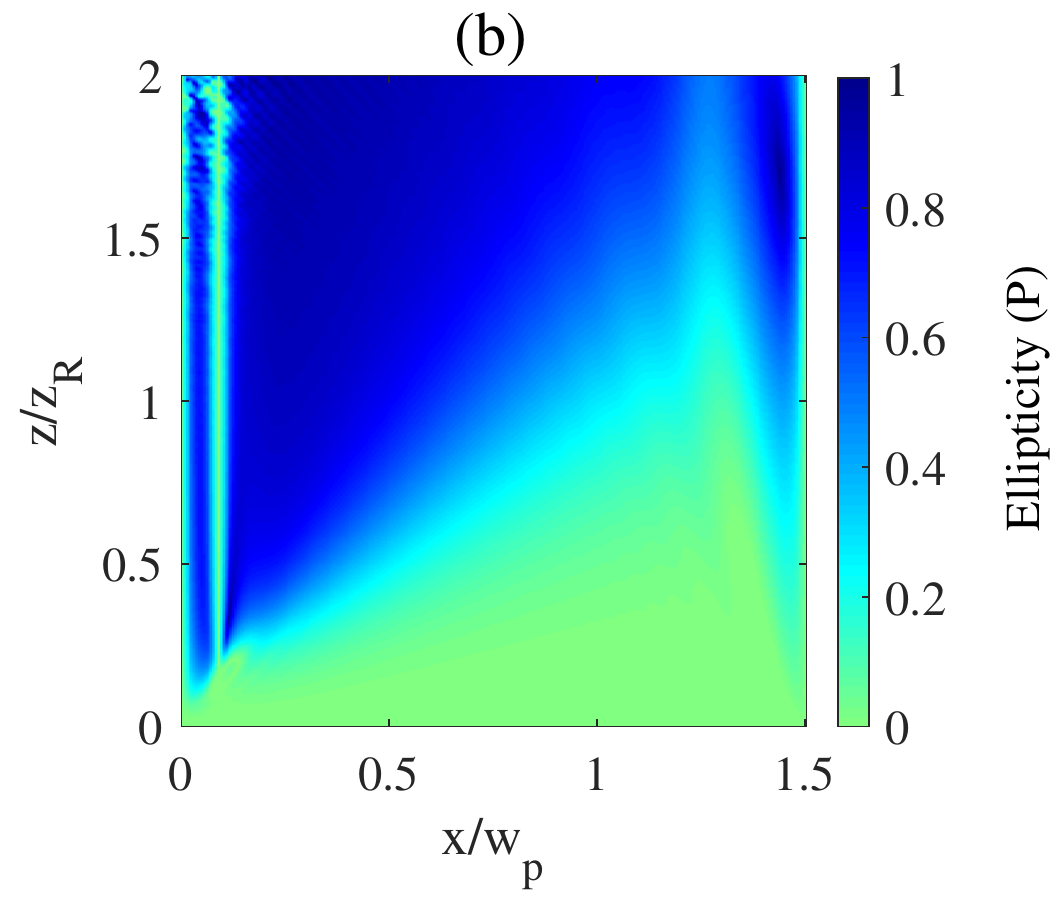}}
\caption{(a) The rotation angle $\theta$ and (b) ellipticity $P$, as a function of transverse $x/w_{\rm{P}}$ and longitudinal $z/z_{\rm{R}}$ position, respectively. The darker blue (red) correspond to  a larger the rotation angle $\theta$ in Fig.6(a), and ellipticity $P$ from 0 (linear polarization) to 1 (circular polarization), other parameters are the same as those in Fig.3.}
\end{figure}

The ellipticity ($P$) of the polarization of the probe light can be quantified as $P=(|\Omega_{\rm{P14}}|-|\Omega_{\rm{P13}}|)/(|\Omega_{\rm{P14}}|+|\Omega_{\rm{P13}}|)$, and its evolution is shown in Fig.6(b). We see a similar spatial behavior of the ellipticity to that of the rotation angle in Fig~.6(a). This provides further evidence that diffraction plays an important role in the evolution of the ellipticity. At short propagation distances, $z<\frac{z_{\rm{R}}}{4}<1$mm,
the two circular components of the probe have not separated yet, so that ellipticity is negligible, while at larger distances, the separation of the two circular components, and hence the ellipticity, increases.

Here we have reported the simplest cases of control beams with identical intensity profiles, but opposite topological charges.  More generally, \textit{e.g.}~for control beams composed of LG beams with different $|l|$ values, or different waists, the linear susceptibilities can take more complicated shapes, and in general the absorption of right and left circular light will differ. In this case, the medium will induce spatially varying circular dichroism as well as birefringence and diffraction.

\section{Conclusion}
In summary, we have shown, by numerical solution of propagation and optical Bloch equations, how polarization information can be transferred from a strong control to a weak probe beam in a double-$V$ atomic level structure. This initially manifests as the generation of a polarization pattern in the probe light. Although not immediately apparent in the intensity profile, the spatial optical coherence is revealed on propagation, where novel patterns expose the internal interaction. We interpret this in the context of differential diffraction or right and left polarized light, where each optical component evolves independently.
We may also consider the effect from the perspective of the optical fields: the right and left circular polarization components of the control light do not interfere, but instead add to a spatially varying polarization structure across the beam profile.  The atomic medium however enables quantum interference due to the nonlinear processes, generating a spatially varying susceptibility. A probe beam interacting with the atomic medium will be modified due to this light-induced susceptibility, affecting the differential phase, amplitude and propagation direction of the two circular polarization components.  Throughout, we show that the susceptibility induced by a polarization structured control field is the source of these subsequent phenomena.  As such, we have demonstrated the generation of controllable spatially dependent birefringence and diffraction.

\section*{Funding}
National Natural Science Foundation of China (NSFC) (62075130).

\section*{Acknowledgments}
SF-A acknowledge financial support from the European Training Network ColOpt, funded by the European Union Horizon 2020 program under the Marie Sklodowska-Curie Action, Grant Agree-ment No.721465. TWC acknowledges support by the National Research, Development and Innovation Office of Hungary (NKFIH) within the Quantum Technology National Excellence Program (Project No. 2017-1.2.1-NKP-2017-00001). JW acknowledges support by the China Scholarship Council (CSC) (No.201906280228). CF acknowledges support for the visiting reseach in Optics Group, Glasgow university, and is grateful to Sonja Franke-Arnold, Thomas W. Clark, and Jinwen Wang for the useful discussions.

\section*{Disclosures}
The authors declare no conflicts of interest.

\section*{Data availability}
Data underlying the results presented in this paper are not publicly available at this time but may be obtained from the authors upon reasonable request.

\end{document}